# Reply to the Comment on "A self-assembled three-dimensional cloak in the visible" in Scientific Reports 3, 2328

Stefan Mühlig[†], Alastair Cunningham[‡], José Dintinger[¶], Mohamed Farhat[†,§], Shakeeb Bin Hasan[†], Toralf Scharf[¶], Thomas Bürgi[‡], Falk Lederer[†], and Carsten Rockstuhl[†]

[†] Friedrich-Schiller-Universität Jena, 07743 Jena, Germany, Institute of Condensed Matter Theory and Solid State Optics, Abbe Center of Photonics, Friedrich-Schiller-Universität Jena, D-07743 Jena, Germany
[‡] Département de Chimie Physique, Université de Genève, Quai Ernest-Ansermet 30, CH-1211 Genève 4, Switzerland
[¶] Optics & Photonics Technology Laboratory, Ecole Polytechnique Fédérale de Lausanne (EPFL), Neuchâtel, CH-2000, Switzerland
[§] Division of Computer, Electrical, and Mathematical Sciences and Engineering, King Abdullah University of Science and Technology (KAUST), Thuwal, 23955-6900, Saudi Arabia

In a recent paper[1] we presented the design, fabrication, and characterization of an optical cloak that renders a small dielectric sphere invisible to an external observer by suppressing its scattered field. The structure has been fabricated by self-assembly techniques and consists of a shell of silver nanoparticles that decorate the dielectric core object. The structure has been devised in the context of the scattering cancelation technique[2,3,4,5,6]. In this technique, an optically small object is covered by a suitable shell to suppress the scattered field. This is achieved in the electric dipolar limit with a shell that scatters the incident light with equal amplitude but π out of phase with respect to the light scattered from the core object. The suppression of scattered light has been proven experimentally and good agreement was found with simulations.

In a recent comment[7] Miller *et al.* pointed out that the presented scheme suffers from enhanced absorption. They claimed that the extinction of the cloaked object becomes larger than that of the bare one in the entire spectral domain, which would disqualify the terminology of a cloak.

We concur with the crux of the argument. The shell of metallic nanoparticles introduces parasitic absorption that becomes larger than the reduction in scattering. Therefore, the extinction of the cloaked object is larger than that of the uncloaked one. However, it is important to stress, and it was appreciated by Miller *et al.*[7], that we never claimed nor suggested a reduction in extinction. The cloak is designed in the context of scattering cancellation technique, which aims at canceling the scattered light only. We write in Ref. 1 explicitly *"We define the scattering efficiency* [the quantity which we discuss throughout the manuscript] *as the ratio of scattered light of the object to be cloaked and of the bare object. Small scattering efficiencies account for highly suppressed scattered fields and therefore a cloaking of the object, i.e. the silica sphere."* If extinction would have been canceled as well, the technique would have been called an extinction cancellation technique.

The issue seems to be whether the structure should be called a cloak or not. Whereas we agree with the statement by Miller *et al.* in Ref. 7 that *"An object creating a large shadow is generally not considered to be cloaked"*, it remains difficult for us to appreciate that an object whose scattering response is restricted to an electric dipole moment may cause anything what could be called a shadow. And just as Miller *et al.* actually write in a nice bon mot in Ref. 7, *"Terminology is usually imprecise."*. For us the appearance of a shadow seems to be restricted to more macroscopic objects which, of course, cannot be cloaked with the technique we exploit. The question remains why we have chosen the terminology of a cloak for a structure that shows a suppressed scattering efficiency?

Guided by ideas published in context of the scattering cancelation technique[2,3,4,5,6], we are convinced that an object is cloaked if it is not perceived by an external observer. We believe that optically small particles are difficult to perceive in extinction. By contrast, if observed in a scattering configuration they can be usually easily seen. Therefore, to obey this definition of a cloak, scattering needs to be reduced.

If the perception of optically small particles in extinction would be technically feasible, the introduction of dark-field microscopy, for example, would not have been required. Eventually, the problem is that the amount of extinct energy of optically small particles, in most cases, remains minor when compared to the energy of the illumination. Therefore, this very small if not to say negligible quantity is required to be detected on top of a huge background signal. Such a signal remains complicated to measure in most experimental schemes since it is vulnerable against noise; although it is not impossible to measure extinction[8]. But the tremendous success of dark-field microscopy can be explained by the technological advantage to measure only a small signal on top of a dark background to see the particles. And this small signal is the amount of scattered light[9].

Therefore, we wish to thank Miller *et al.* for providing this comment since it avoids the misperception of our structure; and potentially others presented in the context of scattering cancelation cloaks. These devices only suppress the amount of scattered light as their name suggests. Extinction, at least in the structures we have studied thus far, is on the contrary enhanced. If a definition of what is understood as a cloak requires the suppression of both scattering and extinction, the structure we presented in Ref. 1 indeed does not deserve to be called a cloak. However, we were guided by the idea that small particles relevant to our work are primarily seen in scattering, which is why employing the terminology of cloak to our structure appeared justified when we wrote the manuscript.